\documentclass[10pt,a4paper]{article}
\usepackage{amsmath,latexsym,amssymb,amsfonts}
\usepackage[dvips]{graphicx}
\usepackage{bm}

\begin{document}

\title{\textbf{Is the firewall consistent?}\\ \large{\textsf{Gedanken experiments on black hole complementarity and firewall proposal}}}
\author{\textsc{Dong-il Hwang}\footnote{dongil.j.hwang@gmail.com},\; \textsc{Bum-Hoon Lee}\footnote{bhl@sogang.ac.kr}\; and \textsc{Dong-han Yeom}\footnote{innocent.yeom@gmail.com}\\
\textit{\small{Center for Quantum Spacetime, Sogang University, Seoul 121-742, Republic of Korea}}}
\maketitle

\begin{abstract}
In this paper, we discuss the black hole complementarity and the firewall proposal at length. Black hole complementarity is inevitable if we assume the following five things: unitarity, entropy-area formula, existence of an information observer, semi-classical quantum field theory for an asymptotic observer, and the general relativity for an in-falling observer. However, large $N$ rescaling and the AMPS argument show that black hole complementarity is inconsistent. To salvage the basic philosophy of the black hole complementarity, AMPS introduced a firewall around the horizon. According to large $N$ rescaling, the firewall should be located close to the apparent horizon.

We investigate the consistency of the firewall with the two critical conditions: the firewall should be near the \textit{time-like} apparent horizon and it should not affect the future infinity. Concerning this, we have introduced a gravitational collapse with a false vacuum lump which can generate a spacetime structure with disconnected apparent horizons. This reveals a situation that there is a firewall outside of the event horizon, while the apparent horizon is absent. Therefore, the firewall, if it exists, not only does modify the general relativity for an in-falling observer, but also modify the semi-classical quantum field theory for an asymptotic observer.
\end{abstract}

\newpage

\tableofcontents

\newpage

\section{Introduction}

The black hole information loss problem \cite{Hawking:1976ra} is a profoundly important issue in quantum gravity. From classical \cite{Israel:1967wq} and semi-classical \cite{Hawking:1974sw} analysis, we already understand that any stationary black hole is characterized by the following three information: mass $M$, charge $Q$ and angular momentum $J$. After a black hole evaporates, what happens to the quantum information? If it cannot be aptly captured with the aid of Hawking radiation, then we immediately see a violation of unitarity and lose fundamental predictability of the theory. Otherwise, how can we be so sure regarding this and how can we restore the information?

After the advent of the AdS/CFT correspondence \cite{Maldacena:1997re}, people became confident about the unitarity of black hole physics, since bulk gravitational dynamics corresponds to the boundary conformal field theory and the boundary conformal field theory should be unitary. However, the question that still crops up in our mind is how does Hawking radiation contains information and is it consistent?

In fact, even before the discovery of AdS/CFT, some people began to consider the consistency of unitarity. Especially, the works by Stephens, t'Hooft and Whiting \cite{Stephens:1993an} and Susskind, Thorlacius and Uglum \cite{Susskind:1993if} contributed to this problem immensely and is respectively known in literature as the \textit{holographic principle} and \textit{black hole complementarity}. According to the black hole complementarity principle, it is customary to think that an asymptotic observer and an in-falling observer of a black hole should satisfy natural laws. Then the semi-classical and unitary quantum field theory should be a good description for the asymptotic observer, while general relativity should serve as a good prescription for the in-falling observer. But, it seems contradictory, since both observers maintain their information and hence the information seem to have copied: one is inside the event horizon, while the other is outside it. However, the black hole complementarity and the natural laws still hold for all observers and are consistent, since two observers cannot communicate among themselves \cite{Susskind:1993mu}. Therefore, although a black hole violates a natural law (the no-cloning theorem), it remains unnoticed like an orderly crime without any witness -- a perfect crime.

However, recently, people started asking questions on the very consistency of the black hole complementarity principle itself. The duplication of information can be observed in case of regular black holes \cite{Yeom:2008qw} or charged black holes \cite{Hong:2008mw,Hong:2008ga,Ge:2005bn}, if we assume large number of scalar fields that contribute to the Hawking radiation. Moreover, it was shown that with  large number of scalar fields, the black hole complementarity can be violated even for a Schwarzschild black hole \cite{Yeom:2009zp}. The number of scalar fields required can be reduced to a reasonable one, if we consider the scrambling time \cite{Hayden:2007cs}. The asymptotic observer and the in-falling observer can communicate with each other inside the black hole and hence the black hole complementarity seems to be inconsistent.

Furthermore, in a recent work, Almheiri, Marolf, Polchinski and Sully (AMPS) \cite{Almheiri:2012rt} have discussed for the inconsistency of black hole complementarity from a different ground. They were able to show that a quantum state, that satisfies classical general relativity for an in-falling observer and that satisfies the unitary quantum field theory for an asymptotic observer, cannot be consistent at the same time. Therefore, it seems that \textit{black hole complementarity is inconsistent not only inside, but also outside the black hole}. To be in live with the original philosophy of black hole complementarity, AMPS suggested the \textit{firewall proposal}. There are some interesting controversy regarding the firewall proposal in the literature \cite{Bousso:2012as,Nomura:2012sw,Mathur:2012jk,Chowdhury:2012vd,Susskind:2012rm,Bena:2012zi,Banks:2012nn,Ori:2012jx}.

In this context, we suggest an interesting toy model for gedanken experiments. We consider a gravitational collapse with a false vacuum lump. This particular example draws inspiration from regular black hole models \cite{Yeom:2008qw,regular}, although it does not necessarily be regular (free from singularity) for our purposes. This model is of interest, since the singularity and horizon structures are non-trivial. We want to address questions like how to define the duplication experiment, how to define the firewall, and whether the firewall can rescue the black hole complementarity prinicple even for this complicated case.

In Section~\ref{sec:black}, we present a concise summary of the black hole information loss problem, motivations and assumptions guiding black hole complementarity and the duplication experiment. In addition, we discuss two important inconsistency arguments for black hole complementarity: large $N$ rescaling \cite{Yeom:2009zp} and the AMPS argument \cite{Almheiri:2012rt}. In Section~\ref{sec:grav}, we discuss the gravitational collapse with a false vacuum lump, using the double-null numerical simulations \cite{Hong:2008mw,doublenull,Hansen:2009kn,Hwang:2010aj}. We analyze the details of the causal structure and discuss some thought experiments relating to black hole complementarity and the firewall proposal. Finally, in Section~\ref{sec:dis}, we summarize and interpret our results.

\section{\label{sec:black}Black hole information loss problem}

In this section, we first discuss \textit{why} does people argue for black hole complementarity. This is related to the analysis of black hole entropy and information. We clarify all the assumptions of black hole complementarity and the consistency check via the duplication experiment. Second, we discuss two counter arguments for black hole complementarity: large $N$ rescaling and the AMPS argument. In addition, we discuss the resolution of the AMPS proposed, so-called firewalls and summarize the recent status of the subject.

\subsection{Why black hole complementarity?}

\subsubsection{Entropy of black holes}

The most remarkable issue in the information loss problem is the entropy of black holes. From the classical point of view, a black hole obeys laws of thermodynamics. From the first law of black hole thermodynamics \cite{Israel:1967wq} and the area law \cite{Hawking:1971vc}, Bekenstein thought that the horizon area is proportional to the thermal entropy of a black hole \cite{Bekenstein:1973ur}. The temperature formula results from the quantum effects around the horizon \cite{Hawking:1974sw}. This entropy is thermal entropy, because for the computation temperature was calculated first and then the entropy was defined as $dS_{\mathrm{th}} = dQ/T$, where $dQ$ is the difference of heat. The natural question to ask is whether this thermal entropy can also be treated as the \textit{statistical} entropy, $S_{\mathrm{st}} = \log {\Omega}$, where $\Omega$ is the number of accessible states.

String theorists believe that the entropy is not only of thermal nature but also has statistical origin. Some string theorists have found the dual of a black hole using D-brane combinations \cite{Callan:1996dv}. It is also known that for certain black holes in the extremal limits with supersymmetry, the entropy obtained in the weak coupling limit is the same as that obtained in the strong coupling limit \cite{Strominger:1996sh}. The entropy could be exactly matched with the entropy formula obtained for some other extreme cases.

Therefore, although there is no formal proof of the thermal and statistical entropy relation,
\begin{eqnarray}
\frac{A}{4} = \log \Omega,
\end{eqnarray}
we will accept this master formula and will try to find out the consequences.

\subsubsection{Information emission from black holes}

Let us specify the information emission from a black hole \cite{Page:1993df,Page:1993wv}. Let us consider a system with number of degrees of freedom $m \times n$ and divide it into two subsystems, $A$ (inside region of the black hole) and $B$ (outside region of the black hole), where the number of degrees of freedom of $A$ is $n$ and that of $B$ is $m$. Note that $m$ and $n$ can vary with time, although $m \times n$ should remain conserved. We can think that initially $m=1$ and as time goes on, $n$ decreases and $m$ increases simultaneously to keep $m \times n$ fixed.

Here, the \textit{mutual information} contained in $A$ and $B$, that is, the information that both $B$ and $A$ share, or in other words, information of $A$ that can be seen by $B$ is $I(B:A)=S(B)-S(B|A)$, where $S(B)=\log m$ is the statistical entropy of $B$ and $S(B|A)$ is the entanglement entropy that is defined by the formula:
\begin{eqnarray}
\rho_{B} &\equiv& \mathrm{tr}_{A} \rho,\\
S(B|A) &=& - \mathrm{tr} \rho_{B} \log \rho_{B},
\end{eqnarray}
where $\rho$ is the density matrix of the total system. In many contexts, people call $S(A)$ or $S(B)$ the \textit{coarse-grained entropy} of $A$ and $B$, while $S(A|B)$ or $S(B|A)$ are known as the \textit{fine-grained entropy} between $A$ and $B$ \cite{Lloyd:1988cn}.

\begin{figure}
\begin{center}
\includegraphics[scale=0.8]{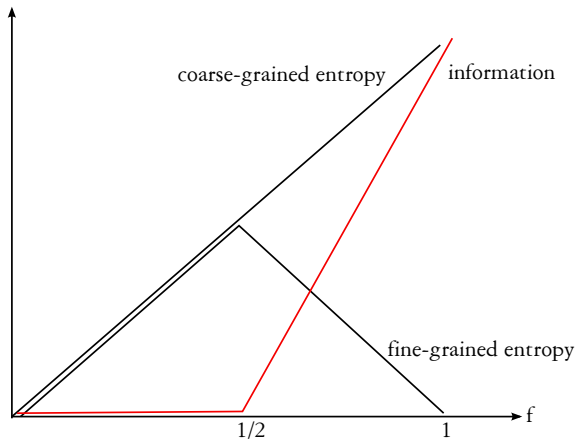}
\caption{\label{fig:information_retention}Emission of information, where $f$ is the ratio of the escaped coarse-grained entropy to the original coarse-grained entropy.}
\end{center}
\end{figure}

We can further proceed by assuming that the system under consideration is pure and random. Page conjectured the following formula in \cite{Page:1993df} and afterwards it was proven in \cite{Sen:1996ph}: if $1 \ll m \leq n$, then
\begin{eqnarray}
S(B|A) &=& \sum_{k=n+1}^{mn} \frac{1}{k} - \frac{m-1}{2n}\\
&\cong& \log m - \frac{m}{2n}.
\end{eqnarray}
Initially, the information emitted is $\cong m/2n$, and therefore is negligible.
If $m > n$, since $S(B|A)=S(A|B)$ for a pure state, one gets
\begin{eqnarray}
S(B|A) &=& \sum_{k=m+1}^{mn} \frac{1}{k} - \frac{n-1}{2m}\\
&\cong& \log n - \frac{n}{2m}.
\end{eqnarray}
Thus, after $n$ becomes greater than $m$, the information emitted is given by $\cong \log m - \log n + n/2m$, and then it gradually increases (Figure~\ref{fig:information_retention}).

In conclusion, the system $A$ begins to emit information to $B$ when its coarse-grained entropy decreases to the half value ($m=n$). Before this time, emitted particles may not contain sufficient information. However, after that time, the original information cannot be compressed within $A$ only and the information possessed by $A$ has to be transferred to $B$ by means of the emitted particles.

\subsubsection{Assumptions}

Let us assume the following:
\begin{description}
\item[Assumption~$1$. Unitarity:] The black hole dynamics is unitary for an asymptotic observer.
\item[Assumption~$2$. Entropy:] $A/4 = \log \Omega$, where $A$ is the area of the black hole and $\Omega$ is the number of accessible states.
\item[Assumption~$3$. Existence of an observer:] There is an observer who can read off information from the black hole.
\end{description}
We further assume the reliability of local quantum field theory and general relativity as methodological tools adopted:
\begin{description}
\item[Assumption~$4$. For asymptotic observer:] The semi-classical method is a reliable description for an asymptotic observer.
\item[Assumption~$5$. For in-falling observer:] General relativity is a good prescription for an in-falling observer.
\end{description}

If we assume the results of the previous two subsections (Assumption~$1$ and Assumption~$2$) so that $A/4 = \log \Omega$ and a black hole begins to emit information when $\log \Omega \rightarrow (1/2) \log \Omega$, then we conclude that the black hole begins to emit information when the \textit{area} of the black hole decreases to half of its initial area. This time scale is of the order of the lifetime of a black hole $\sim M^{3}$ and is called the \textit{information retention time} \cite{Susskind:1993mu}. There are many There are many instances when the black hole can be treated in a semi-classical way, i.e., although the area of the black hole decreased to half of its initial value, the black hole is still large enough. Then, the only way to take out information from the large black hole is with the aid of Hawking radiation. Therefore, \textit{information should be emitted through Hawking radiation}.

\subsubsection{Duplication experiment and black hole complementarity}

\begin{figure}
\begin{center}
\includegraphics[scale=0.8]{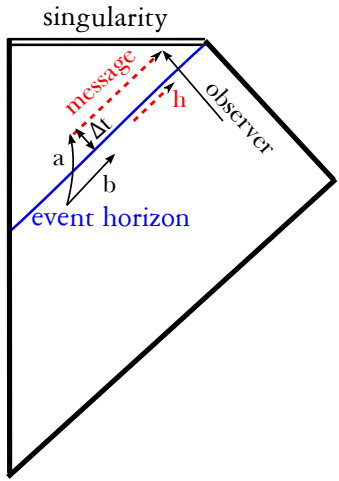}
\caption{\label{fig:Schwarzschild_duplication}The duplication experiment. $a$ and $b$ are a pair of correlated spins. The observer sees $h$ which is a copy of $a$, after the information retention time via Hawking radiation. $a$ should be sent to the out-going direction after the time $\Delta t$ in order to be observed. If the observer sees both $a$ and $h$, since they are both correlated to $b$, it violates the no-cloning theorem and unitarity.}
\end{center}
\end{figure}

Let us think of a specific situation (Figure~\ref{fig:Schwarzschild_duplication}) \cite{Susskind:1993mu} and consider a series of experiments in which a pair of correlated spins are simulated outside the event horizon. One of the pair that falls into the black hole is denoted by $a$ and the other pair that remains outside the black hole is $b$. If Hawking radiation contains information, then information about $a$ can be emitted through Hawking radiation and we call it $h$. According to Assumption~$3$, if there is an observer who can measure the state of $h$, falls into the black hole, and measures the state of $a$, eventually we will know that the collected information $a$ and $h$ are both correlated to $b$. This implies that the observer sees a duplication of states, which is no allowed by quantum mechanics. We will call this type of experiment a \textit{duplication experiment}.

Susskind and Thorlacius \cite{Susskind:1993mu} were able to answer several important questions related to the duplication experiment. If the observer sees both $a$ and $h$, he/she has to wait until the completion of the information retention time. However, if the original free-falling information $a$ reaches the singularity of the black hole, then there is no chance to see the duplication. In order to be able to see the duplication, the free-falling information $a$ should be sent to the out-going direction during the time interval $\Delta t$.

We can estimate the time interval $\Delta t$ in the Schwarzschild space-time:
\begin{eqnarray}
ds^{2} = - \left( 1- \frac{2M}{r} \right) dt^{2} + \left( 1- \frac{2M}{r} \right)^{-1}
dr^{2} + r^{2} d\Omega^{2}.
\end{eqnarray}
The horizon is located at $r_{\mathrm{h}} = 2M$ and the Hawking temperature $T$ is of the order $\sim 1/M$. Therefore, the lifetime $\tau \sim M^{3}$.

For the next calculation, we will begin by making a comment on a simple extension to Kruskal-Szekeres coordinates \cite{Susskind:1993mu,Lowe:2006xm}. We can neglect the angular part without loss of generality and assume the form of the metric to be
\begin{eqnarray}
ds^{2} = F(R) \left( -R^{2} d\omega^{2} + dR^{2} \right).
\end{eqnarray}
To compare with the original metric, the following definitions are made:
\begin{eqnarray}
d \omega^{2} &=& \frac{dt^{2}}{r_{\mathrm{h}}^{2}},  \\
R^{2}F(R) &=& r_{\mathrm{h}}^{2} \left( 1- \frac{2M}{r} \right),  \\
F(R)dR^{2} &=& \left( 1- \frac{2M}{r} \right)^{-1} dr^{2}.
\end{eqnarray}
In terms of the coordinate $R$, the singularity occurs at $R^{2} = - r_{\mathrm{h}}^{2}$; and the horizon is located at $R=0$. Now, we can choose another metric and coordinate $(U,V)$ as follows:
\begin{eqnarray}
V &=& R e^{\omega},  \\
U &=& - R e^{-\omega},  \\
ds^{2} &=& -F(R) dU dV.
\end{eqnarray}
Here, the singularity is located at $UV = r_{\mathrm{h}}^{2}$.

We can restate the condition for a duplication experiment in a Schwarzschild black hole.
The first observer falls into a black hole and sends a signal to the out-going direction in time interval $\Delta t$.
Now assume that a second observer hovers around the horizon at a distance of the order of the Planck length ($r \sim 2M + l_{\mathrm{Pl}}$) and jumps into the black hole at the information retention time which is of the order of $\sim \tau$.
Then, the initial location of the second observer is $V=R e^{\omega}$,
where $R \equiv R_{0}$ and $\omega \sim \tau / r_{\mathrm{h}}$.\footnote{Here, $R_{0}$ is a length scale that is relevant for the Kruskal-Szekeres coordinates. For a precise calculation of $R_{0}$, we need to calculate $R_{0}$ from $r \sim 2M + l_{\mathrm{Pl}}$ and $t \sim M^{3}$. We will not do the detailed calculation, but it is sufficient to notice that $M \gg R_{0} \gtrsim l_{\mathrm{Pl}}$.}
Before reaching the singularity, the second observer will spend time (in terms of $U$) around $\sim r_{\mathrm{h}}^{2} / V$
since the singularity is reached where $UV = r_{\mathrm{h}}^{2}$.
Therefore, the first observer should send a signal around the time $\Delta t \sim e^{- \tau / r_{\mathrm{h}}}$.
Hence, the duplication may be possible if one can send a signal between the time
\begin{eqnarray}\label{eq:cons}
\Delta t \sim \frac{M^{2}}{R_{0}} \exp{-\frac{\tau}{r_{\mathrm{h}}}},
\end{eqnarray}
where $\tau$ is the information retention time.

Then, to send a quantum bit during the time interval $\Delta t$, it has to satisfy the uncertainty relation $\Delta t \Delta E \gtrsim 1$. The required energy to send a quantum bit of information in $\Delta t$ is $\sim \exp M^{2}$, which is greater than the original mass of the black hole $M$. Therefore, the duplication experiment seems to be improbable in real situations \cite{Susskind:1993mu}.

According to Susskind and Thorlacius, although information is duplicated, there seems to be no problem if no observer can see the violation of the natural laws. In other words, there is no global description both for an in-falling observer and an asymptotic observer at the same time. We have to choose one of them. In this sense, two observers are complementary. This principle is known as \textit{black hole complementarity} or observer complementarity \cite{Susskind:1993if}.

Black hole complementarity is consistent with two paradigms: the membrane paradigm \cite{Thorne:1986iy} and the D-brane picture \cite{Callan:1996dv}. In membrane paradigm, a black hole has a membrane around the event horizon, the so-called stretched horizon. If we send an object into a black hole, the object is stretched and scrambled on the horizon. The outside observer cannot see the object disappearing beyond the horizon. Therefore, for an outside observer, information is located on the horizon and eventually escapes from the black hole via Hawking radiation. The scrambling occurs in the following order of time:
\begin{eqnarray}
\tau_{\mathrm{scr}} \sim \beta M \log M,
\end{eqnarray}
and this is called the \textit{scrambling time}. In this paper, we write a coefficient factor $\beta$ explicitly for further clarity, where it should be order one and can vary in realistic situations, since the scrambling is a statistical behavior. According to Hayden and Preskill \cite{Hayden:2007cs}, after a black hole approaches the information retention time, if one sends small bits of information, this will quickly escape from the black hole after the scrambling time. Note that although we consider the scrambling time, the consistency relation still holds: from Equation~(\ref{eq:cons}), we find that in order to see the duplication, $\Delta E \sim M > M$. Therefore, people believed that \textit{black hole complementarity is marginally true, even with the scrambling time}.

\subsection{Inconsistency of `old' black hole complementarity}

Now we will introduce two important arguments against the original version of black hole complementarity. One is the so-called large $N$ rescaling \cite{Yeom:2009zp} and the other is the AMPS argument \cite{Almheiri:2012rt}.

\subsubsection{Large $N$ rescaling}

Let us assume that $G=c=1$ and $\hbar$ remains explicitly. Then, length, mass, and time dimensions are the same. In this subsection, we will change the number of massless scalar fields $N$ and hence we will scale the strength of the Hawking radiation. We assume that there is one scalar field $\phi$ that contributes to the formation of a black hole; the other $N$ number of fields are not used to form the black hole, and they only contribute to the Hawking radiation.

First, let us assume $N=1$. Then the semi-classical equations of motions (up to order $\hbar$) are as follows:
\begin{eqnarray} \label{semi-classical}
G_{\mu \nu} &=& 8 \pi (T_{\mu \nu} + \hbar \langle T_{\mu \nu} \rangle), \\
\phi_{;ab}g^{ab} &=& 0,
\end{eqnarray}
where $\phi$ is a scalar field used to form a black hole.

Now we define the re-scaling with the following rule: if a quantity $X$ which does not explicitly depend on $\hbar$ has a dimension $[X]=L^{\alpha}$ with a certain number $\alpha$, we define a rescaled $X'$ as
\begin{eqnarray} \label{eq:res}
X' = \sqrt{N^{\alpha}} X.
\end{eqnarray}
Then, we claim that after rescaling \textit{all} possible quantities, they are solutions of the following equation:
\begin{eqnarray} \label{semi-classical_2}
G'_{\mu \nu} &=& 8 \pi (T'_{\mu \nu} + N \hbar \langle T'_{\mu \nu} \rangle),\\
\phi'_{;ab}g'^{ab} &=& 0.
\end{eqnarray}
This is easy to check: $G_{\mu \nu}$ has a dimension of $L^{-2}$, $T_{\mu \nu}$ also has a dimension of $L^{-2}$, and $\langle T_{\mu \nu} \rangle$ has a dimension $L^{-4}$ in the one-loop order. Hence, $G'_{\mu \nu} = G_{\mu \nu}/N$, $T'_{\mu \nu} = T_{\mu \nu}/N$, and $\langle T'_{\mu \nu} \rangle = \langle T_{\mu \nu} \rangle / N^{2}$. Then,
\begin{eqnarray} \label{semi-classical_3}
G_{\mu \nu} = N G'_{\mu \nu} = 8 \pi (T_{\mu \nu} + \hbar \langle T_{\mu \nu} \rangle) = 8 \pi (N T'_{\mu \nu} + \hbar N^{2} \langle T'_{\mu \nu} \rangle),
\end{eqnarray}
and thus our claim is correct. Also, it is easy to check the same relation for the Klein-Gordon equation for a scalar field $\phi$.

In conclusion, for any given quantities which are solutions of Equation~(\ref{semi-classical}), the corresponding rescaled quantities are the solutions of Equation~(\ref{semi-classical_2}) with $N$ massless fields.
Three important remarks regarding the large $N$ rescaling are as follows.
\begin{description}
\item[Conformal invariance of the causal structure:] The rescaling conserves the very causal structure of the metric since it scales the unit length and the unit time in the same way. Therefore, we can use the same Penrose diagram for the $N = 1$ case.
\item[Semi-classicality:] If we can simulate a sufficiently large $N$ universe, even if a region has a large curvature in the $N = 1$ case (in Planck units), we can find a universe where the curvature can be rescaled to a sufficiently smaller value (in Planck units). Therefore, large $N$ rescaling makes results trustable in the semi-classical sense.
\item[Generalization to other matter fields:] We can generalize to include more complicated matter fields: e.g., complex scalar field, complicated potential, etc. For these cases, we have to rescale coupling constants when we vary the number of scalar fields. As long as the coupling constants are free parameters of the theory, it is always allowed in principle.
\end{description}

Let us apply the large $N$ rescaling to the information retention time and the scrambling time.
\begin{description}
\item[-- Information retention time:] We rescale all length, mass, and time parameters by $\sqrt{N}$. Now, the information retention time $\tau$ for mass $M$ and the single scalar field is rescaled to $\tau'$ for mass $M'=\sqrt{N}M$ where $N$ is a certain number and
\begin{eqnarray}
\tau &\sim& M^{3},\\
\tau' &\sim& \frac{M'^{3}}{N} = \frac{(\sqrt{N}M)^{3}}{N} = \sqrt{N}M^{3}.
\end{eqnarray}
Now, we have to divide the lifetime by $N$, since there are $N$-independent fields that contribute to the Hawking radiation. Note that the size $r_{\mathrm{h}}$ will be rescaled to $r_{\mathrm{h}}' = \sqrt{N}r_{\mathrm{h}}$. Therefore, under large $N$ rescaling, the ratio between the temporal size and the spatial size remains invariant:
\begin{eqnarray}
\frac{\tau}{r_{\mathrm{h}}} = \frac{\tau'}{r_{\mathrm{h}}'}.
\end{eqnarray}
Note that, although $\tau/r_{\mathrm{h}}$ is invariant under the large $N$ rescaling, each conformally equivalent distances should be stretched with $\sqrt{N}$ factor. Therefore, in general, in $N=1$ limit, the duplication may be observed if one can send a signal between the time interval $\Delta t \sim (M^{2}/R_{0}) \exp{-\tau/r_{\mathrm{h}}}$, where $\tau$ is the information retention time ($\sim M^{3}$). On the other hand, in the large $N$ rescaled case,
\begin{eqnarray}
\Delta t' \sim \sqrt{N} \left(\frac{M^{2}}{R_{0}} \right) \exp{-\frac{\tau'}{r'_{\mathrm{h}}}} \sim \sqrt{N} \Delta t.
\end{eqnarray}
Note that $M^{2}/R_{0}$ has a length dimension and hence we add only $\sqrt{N}$ for the rescaling.
From the uncertainty relation, the energy required is
\begin{eqnarray}
\Delta E' \sim \frac{1}{\sqrt{N}} \left(\frac{R_{0}}{M^{2}}\right) \exp{\frac{\tau}{M}},
\end{eqnarray}
and since the consistency of complementarity requires $\Delta E' > M' = \sqrt{N} M$,
the consistency condition turns out to be
\begin{eqnarray}
\exp{M^{2}} > N \frac{M^{3}}{R_{0}}.
\end{eqnarray}
This condition can be violated by assuming to be sufficiently large of the order $N \sim \exp M^{2}$ \cite{Yeom:2008qw,Hong:2008mw,Hong:2008ga}.

\item[-- Scrambling time:] The scrambling time is $\sim M \log M \sim M \log S$, where $S$ is the entropy of the black hole \cite{Hayden:2007cs}. In fact, the rescaling is done for $M \log S/\hbar$, and it turns out to be $\sim \sqrt{N} M \log \sqrt{N}M$. Thus, in a large $N$ universe, the time scale becomes
\begin{eqnarray}
\Delta t' \sim \sqrt{N} \left(\frac{M^{2}}{R_{0}}\right) \exp{\left(-\beta\frac{M \log \sqrt{N}M}{M}\right)} \sim \sqrt{N} \left(\frac{M^{2}}{R_{0}}\right) \exp{\left(-\beta\log \sqrt{N}M\right)}.
\end{eqnarray}
From the uncertainty relation, the energy required is
\begin{eqnarray}
\Delta E' \sim \frac{1}{\sqrt{N}} \left(\frac{R_{0}}{M^{2}} \right) \left(\sqrt{N}M\right)^{\beta}.
\end{eqnarray}
Due to the consistency of complementarity $\Delta E' > \sqrt{N} M$, and hence
the consistency condition becomes
\begin{eqnarray}
\left(\frac{R_{0}}{M^{2}} \right) M^{\beta-1} > \sqrt{N} N^{(1-\beta)/2}.
\end{eqnarray}
As we noted, the scrambling time is the \textit{statistical} notion. After the \textit{order of} the scrambling time, information comes out; but the details of the time can be different case by case. Only the statistical average of the time for various situations follows $\sim \beta M \log M$. Therefore, if this inequality is violated for $\beta = 1$, then this is sufficient to show the inconsistency of black hole complementarity. The condition is
\begin{eqnarray}
\left(\frac{R_{0}}{M^{2}} \right) < \sqrt{N}.
\end{eqnarray}
Here, we can see that this inequality can hold by assuming a sufficiently large $N$ \cite{Yeom:2008qw,Yeom:2009zp}.\footnote{In literature \cite{Hayden:2007cs}, people approximately assume the pre-exponential factor as one; and then the scrambling time seems to be marginal for this inequality. However, here we introduce the pre-exponential factor and large $N$ rescaling at the same time and hence the violation of the consistency condition is clearer.}
\end{description}
In this sense, the black hole complementarity principle can be violated even if we consider a Schwarzschild black hole. The number of scalar fields required can be made sufficiently small, if we consider the scrambling time.

Even in the case when we do not consider the scrambling time and only consider the information retention time, still we find that black hole complementarity is inconsistent. Although the required number of scalar fields is exponentially large, it is still not infinite and hence is allowed by string theory in principle \cite{Dvali:2007hz}. There are some typical doubts due to our misunderstanding of the large $N$ rescaling and these issues can be answered as follows:
\begin{enumerate}
\item People think that large $N$ induces strong Hawking radiation and hence the black hole evaporates too quickly and hence cannot be thought of as a semi-classical object \cite{Dvali:2007hz}. However, this is not true. We not only increase the strength of the Hawking radiation, but also the size of the black hole at the same time. Dvali suggested that a semi-classical black hole should be larger than $\sqrt{N}$: in other words, $M > \sqrt{N}$. In our discussions, $M \gg 1$ and, by the rescaling argument, $M'=\sqrt{N}M \gg \sqrt{N}$ always holds. Therefore, in the large $N$ limit, the back-reactions from Hawking radiation \textit{decrease} and the lifetime of the black hole \textit{increases}.
\item People also worry the higher order quantum corrections while including large number of scalar fields. This is a natural concern, but our focal attention is not on an general gravitational system, but a very special one that can be allowed in principle. So, the real puzzle stems from the fact \textit{whether the large $N$ inevitably require strong quantum correction effects or not}. The answer to the question is `it may not be', in principle. First of all, the curvature corrections will be suppressed due to rescaling. Second, higher loop corrections of the matter fields depend on their couplings. If we assume that $N$ scalar fields are independent of each other, then the higher order terms will be reasonably suppressed, as they are proportional to $N \hbar^{2}$, while this in fact is sufficiently smaller compared to the term $N \hbar$. If this assumption is in principle possible, then it is a good playground to test the consistency of black hole complementarity, unless we find a fundamental limitation of this assumption.
\end{enumerate}

\subsubsection{Duplication experiment outside the event horizon}

\begin{figure}
\begin{center}
\includegraphics[scale=1]{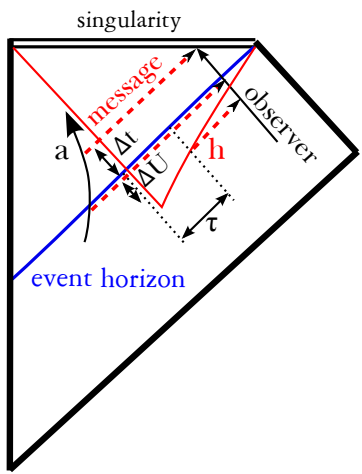}
\caption{\label{fig:Schwarzschild_duplication2}The duplication experiment conducted outside the event horizon. $a$ is in-going information and a signal is being sent to the out-going direction. This can be both inside and outside of the event horizon. To see $a$ inside the event horizon, $a$ should be sent to an out-going direction after time $\Delta t$, while to see it outside the event horizon, $a$ should be sent after time $\Delta U$.}
\end{center}
\end{figure}

Till now, we have ignored the contribution of the subexponential factor of $\Delta t$. Note that for an evaporating black hole, the apparent horizon is located outside the event horizon. Therefore, the duplication experiment can be conducted not only inside but also outside the event horizon. To see the duplication outside the black hole, the in-falling information should be sent to an out-going direction between the new time interval $\Delta U$. In general, we expect $\Delta U \ll \Delta t$, and the difference comes into effect due to the subexponential factor.

Let us discuss the details now. Let us assume that the black hole mass is initially $M_{1}$ and as time goes on, it shrinks to $M_{2}$ after a time scale $\tau$. Now let us define a duplication observer who maintains its radius around $r_{1} = 2M_{1}$ until the time $\tau$ and eventually falls to $r_{2}+ l_{\mathrm{Pl}} \sim 2M_{2}$ (still outside the apparent horizon) along the in-going null direction in time $\tau$.

Then, we should calculate the difference of the coordinate $U$ between the two points $(V\simeq\tau, r=r_{1})$ and $(V\simeq\tau, r\simeq r_{2})$ (Figure~\ref{fig:Schwarzschild_duplication2}). The $\omega$ coordinate is approximately the same: $\omega \sim \tau/r_{2}$ for both the points. In addition, the $R$ coordinates are
\begin{eqnarray}
R = M\left(\frac{r-2M}{2M}\right)^{1/2} \exp{\frac{r}{4M}}
\end{eqnarray}
by using a simple coordinate transformation. Therefore, $\Delta U = \Delta R \exp{-\omega}$ and hence approximately given by
\begin{eqnarray}\label{eq:cons2}
\Delta U \simeq \sqrt{M_{2} \Delta M} \exp{{- \frac{\tau}{r_{2}}}},
\end{eqnarray}
where $\Delta M$ is mass decreased due to the evaporation during time $\tau$.

Note that, in Equation~(\ref{eq:cons}), we have omitted the subexponential factor where the factor is approximately $M_{2}^{2}$. Of course, in principle we can restore it back and compare with the sub-exponential factor of Equation~(\ref{eq:cons2}) and in general the latter is shorter than the former as
\begin{eqnarray}
\frac{\Delta U}{\Delta t} \simeq \frac{\sqrt{M_{2} \Delta M}}{M_{2}^{2}/R_{0}} \ll 1.
\end{eqnarray}
Therefore, it is easier to see the duplication not outside but inside of the black hole. However, the duplication experiment is dominated by the exponential factor and two time scales share the same exponential factor. Hence, $N$ required for a successful duplication experiment is similar for both the cases.

\begin{description}
\item[-- Outside the event horizon considering the information retention time:] Here,
\begin{eqnarray}
\Delta M = M_{1}-M_{2} \simeq \left( \sqrt{2}-1 \right) M_{2}.
\end{eqnarray}
Therefore, using the same argument as that of the previous subsection, the duplication experiment is possible even outside the event horizon, as long as $N \gtrsim \exp M^{2}$ scalar fields are possible to be taken into account.

\item[-- Outside the event horizon considering the scrambling time:] With the scrambling time, we can also define a duplication experiment outside the event horizon. Here, $\Delta M \simeq \tau_{\mathrm{scr}}/M_{2}^{2}$. After large $N$ rescaling and applying the uncertainty relation, the required energy $\Delta E'$ is
\begin{eqnarray}
\Delta E' \sim \frac{1}{\sqrt{\log M_{2}}} \frac{1}{\sqrt{N}} \exp{\log \beta \sqrt{N}M_{2}}.
\end{eqnarray}
The duplication experiment is possible even outside the event horizon, if $\Delta E' < \sqrt{N}M_{2}$ and hence
\begin{eqnarray}
N^{2-\beta} > \frac{M_{2}^{\beta-1}}{\log M_{2}}.
\end{eqnarray}
Again, for $\beta = 1$, we see that the duplication observation requires a reasonable number of scalar fields\footnote{Up to now, we think that the consistency condition is $\Delta E > M$. However, in practice, we cannot use all the energy $M$ to send a signal and there can be a certain limitation: we can use at most $\epsilon M$ amount of energy to send a signal, where $\epsilon < 1$. Then $N$ required becomes $N > (\epsilon \log M_{2})^{-1}$ for outside the horizon (by choosing $\beta=1$). This implies that required $N$ can be greater than one. However, it is also true that such a required number can be still reasonably small.}:
\begin{eqnarray}
N > \frac{1}{\log M_{2}}.
\end{eqnarray}
\end{description}

In conclusion, if there are sufficiently large number of scalar fields, then the duplication experiment is possible not only inside but also outside the event horizon. Therefore, to prevent such a duplication experiment, the in-going observer should be killed very near the \textit{apparent horizon}, rather than the event horizon after a certain time scale (information retention time or scrambling time). This gives us a hunch on the location of the firewall, which we will discuss in the next subsection.

\subsubsection{AMPS argument and the firewall controversy}

Recently, Almheiri, Marolf, Polchinski, and Sully \cite{Almheiri:2012rt} suggested that the assumptions of black hole complementarity are inconsistent using other arguments. Let us define two sets of quantum operators of the matter fields: for the initial state $a$, $a^{\dag}$ and for the final state $b$, $b^{\dag}$. According to Assumption~$5$, the information regarding the collapse will be described by the ground state $a_{\omega}^{\dag}a_{\omega} = 0$. As far as, Assumptions~$1$, $2$ and $3$, after the information retention time, even if we put a small information, that will be emitted through Hawking radiation after the scrambling time. From Assumption~$4$, there is a unitary and semi-classical description between the initial and the final state given by
\begin{eqnarray}
b=\int_{0}^{\infty} \left( B(\omega) a_{\omega} + C(\omega) a_{\omega}^{\dag} \right) d\omega.
\end{eqnarray}
In addition, according to Assumption~$4$, an asymptotic observer can measure particles into an eigenstate of $b^{\dag}b$. Note that the full state cannot be both $a| \Psi \rangle = 0$ and at the same time the eigenstate of $b^{\dag}b$. This is a contradiction.

To resolve this contradiction, they suggested two alternatives. One is to drop Assumption~$5$ so that the in-falling observer watches a sudden change (violation of the equivalence principle); they assumed that there is a \textit{firewall} near the horizon for an in-falling observer. The other is to drop Assumption~$4$ so that the asymptotic observer sees a radical non-local effects.

Here, we briefly summarize the discussions on the firewall proposal:
\begin{description}
\item[It's inconsistent, so what?:] Bousso \cite{Bousso:2012as} argued that AMPS clearly shows the potential inconsistency of the black hole complementarity. However, this is not a problem, since the inconsistency cannot be observed in principle\footnote{Recently, Bousso changed his opinion in the second version of \cite{Bousso:2012as}, while this paragraph follows the first version of the paper.}. To notice the inconsistency, one has to compare the in-falling and the asymptotic observer. However, the communication is impossible. Therefore, although it is apparently inconsistent, we do not have to modify the black hole complementarity. This interpretation is consistent with Nomura, Varela and Weinberg \cite{Nomura:2012sw} and Banks and Fischler \cite{Banks:2012nn}, in the sense that the in-falling observer and the asymptotic observer correspond to different detectors and hence cannot be compared in a naive way. Black hole complementarity is for the whole quantum states, while an in-falling observer and an asymptotic observer are only parts of the whole quantum states. Therefore, the comparison between the two observers are not well-defined.

    These kind of arguments relies heavily on the fact that two observers -- asymptotic and in-falling -- cannot communicate in the semi-classical sense. However, this is not true. Large $N$ rescaling shows that they can communicate with each other even in a semi-classical sense. Therefore, we cannot simply avoid the potential inconsistency and one has to see what happens when two observers meet. In this sense, we think that this opinion cannot really help to unveil the problem of black hole complementarity.

\item[The fuzzball and approximate complementarity:] Mathur and Turton \cite{Mathur:2012jk} and Chowdhury and Puhm \cite{Chowdhury:2012vd} tried to connect the firewall argument with the fuzzball conjecture. If the energy scale of a probe is smaller or similar order with the Hawking temperature, then the probe sees a fuzzball and it is approximately a realization of the firewall. However, if the energy scale of the probe is \textit{much} larger than that of the Hawking temperature, then the probe can free-fall into the black hole; the probe satisfies the equivalence principle. They called this phenomenon \textit{approximate complementarity}.

    However, there still remain some questions. If a free-fall is allowed, then large $N$ rescaling makes the duplication experiment to be possible. Up to now, it seems unclear how does the fuzzball argument prevent this possibility. One important remark is that one can distinguish a fuzzball in the macroscopic scales \cite{Bena:2012zi}. Therefore, if the fuzzball conjecture is in general true, then it not only modifies Assumption~$5$, but also Assumption~$4$.


\item[The firewall is a new singularity:] Susskind \cite{Susskind:2012rm} remarked that if there is a firewall, then it should be regarded as a new type of singularity. The entanglement of a black hole with the Hawking radiation causes the singularity to migrate toward the horizon and eventually intersect it at the information retention time. Therefore, we have to consider the singular horizons -- firewall -- after the information retention time. In addition, after the information retention time, if some information falls into the black hole, then the horizon increases along the space-like direction and eventually becomes singular after the scrambling time.

    This interpretation is worthwhile to discuss further. We will comment on this in the next subsection.

\item[The semi-classical point of view:] Ori \cite{Ori:2012jx} thinks that it is not inconsistent to extend the semi-classical quantum field theory beyond the event horizon. In other words, what he means is that there is no good justification to believe in the existence of a firewall. Therefore, if we accept that there is no firewall and black hole complementarity is not true, then the next possible choice is to consider the regular black hole/remnant picture or the baby universe scenario. This conclusion was also made by the authors \cite{Yeom:2009zp} and probably also by a number of general relativists.
\end{description}

\begin{figure}
\begin{center}
\includegraphics[scale=1]{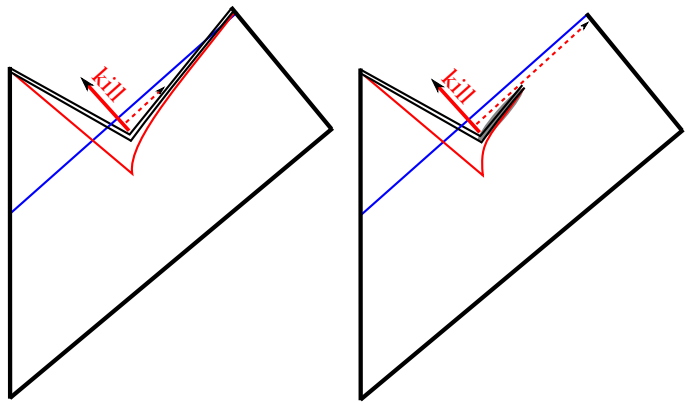}
\caption{\label{fig:wall_3}Left: The signal from the firewall (thick red arrow) kills the in-falling information. However, it does not affect outside as the effects (red dotted arrow) is screened by the apparent horizon. Right: If the apparent horizon is disconnected, then there is no screen such that the effects (red dotted arrow) can modify the asymptotic infinity.}
\end{center}
\end{figure}

\subsubsection{Is the firewall-singularity consistent?}

Let us extend the discussion of \cite{Susskind:2012rm}. We require the following two conditions for a firewall:
\begin{enumerate}
\item It should prevent the duplication experiment.
\item We do not want to modify the macroscopic scale (asymptotic) semi-classical theory.
\end{enumerate}
As we observed in the previous subsections, the duplication experiment can be done outside the event horizon by assuming a reasonable number of scalar fields. Therefore, to prevent the duplication experiment, the firewall should be outside the event horizon. Then, the reasonable place of the firewall will be close to the \textit{time-like} apparent horizon \cite{Ashtekar:2004cn}. However, due to the second condition, the time-like object should not affect the future infinity. Now, the question is whether it is indeed consistent.

Of course, it is natural to think that the time-like firewall should affect the future infinity, since it is outside the event horizon. However, in this paper, \textit{we try to maintain the original motivation of the firewall proposal} to study the consistency of the firewall proposal. Then, what should we assume? If one wants to hold the conservative point of view so that one believes the firewall does not affect the future infinity, then we must require these two properties at the same time: (1) the firewall sends signals along the in-going direction to kill the in-falling information (thick arrow on the left of Figure~\ref{fig:wall_3}) and (2) although there is a bouncing effect along the out-going direction, the \textit{apparent horizon} should work as a screen to any out-going effect from the firewall (dotted arrow on the left of Figure~\ref{fig:wall_3}).

In this paper, we will eventually conclude as follows: even though we further assert the previous two assumptions, we cannot prevent to affect the future infinity. These properties can be falsified if \textit{the apparent horizon is separated} (the right of Figure~\ref{fig:wall_3}). Then the apparent horizon cannot screen out the out-going effects. Or, we can say that, the firewall becomes a naked singularity. In any case, the firewall should affect the future infinity. In the next section, we will realize such separated horizons.

\section{\label{sec:grav}Gravitational collapse with a false vacuum lump}

In this section, we want to discuss a gravitational collapse with a false vacuum lump. This is motivated by regular black hole models, but it does not necessarily have to be regular. We calculate this using the double-null formalism and numerical implementations. After we specify the causal structures, we will discuss \textit{gedanken experiments} on black hole complementarity and the firewall proposal.

\subsection{Model}

\subsubsection{Regular black hole models}

Regular black holes are introduced to explain the problem of singularity arising in black holes. According to the singularity theorem, if we assume the following three things and general relativity, we cannot avoid the existence of a singularity \cite{Hawking:1973uf}: (1) global hyperbolicity, (2) the null energy condition and (3) the existence of a trapped surface.

In order to define a black hole, we cannot avoid the last assumption, the existence of a trapped surface. Therefore, one may choose from the following possibilities \cite{regular}: (1) modify general relativity around the singularity, (2) violate the global hyperbolicity and introduce a Cauchy horizon or (3) violate the null energy condition.
\begin{description}
\item[Modification of general relativity:] Around the singularity, general relativity should be radically modified. One possibility is to consider the spacetime to be fuzzy and not well-defined. The other possibility is to choose a good metric ansatz, although the metric should be affected by some quantum gravitational corrections. However, it crucially depends on the details of the quantum gravity theory.
\item[Violation of the global hyperbolicity:] One may assume that a black hole solution is the same as the Schwarzschild solution for large $r$ (or any known static solution), while there is a non-trivial matter core inside the black hole. In general, the matter core collapses to form a singularity. However, if the formation of the singularity can be postponed and an inner horizon appears, then one may construct a regular solution, without violating the null energy condition. However, these models typically have an inner horizon and these inner horizons are Cauchy horizons in the static limit. Therefore, the global hyperbolicity is violated.

    In general, if there is a Cauchy horizon, mass inflation is inevitable \cite{Poisson:1989zz}. Then, even in the absence of a central strong singularity, a curvature singularity (with non-zero area) can be formed. Therefore, the resolution of mass inflation is again required.
\item[Violation of the null energy condition:] If we assume a certain amount of matter that violates the null energy condition, then it is not quite difficult to find a regular black hole solution, since the formation of a singularity can be postponed by matter fields. The problem is the origin of matter. One may assume phantom matter which is not ruled out by cosmological observations. One trivial example that realizes phantom matter is a ghost field.

    One problem is that a ghost field makes the field theory unstable. Therefore, it is fair to say that observations cannot rule out the existence of phantom/ghost-like matter; also, there is no observational justification in favor of the use of phantom/ghost matter. On the other hand, if a false vacuum bubble can emit negative energy flux, then it can form a negative energy bath \cite{Hansen:2009kn}, although this particular process requires many assumptions.
\end{description}

\subsubsection{Justification of the physical possibility}

In spite of the potential problems, in this paper, we introduce a regular black hole model. Let us first comment on the static solution of Frolov, Markov, and Mukhanov \cite{Frolov:1988vj}.
The metric and the energy-momentum tensor of the shell are as follows:
\begin{eqnarray} \label{metric}
ds^{2} = -\left(1-\frac{2m(r,l)}{r}\right) dt^{2}+\left(1-\frac{2m(r,l)}{r}\right)^{-1}dr^{2}+r^{2}d\Omega^{2},
\end{eqnarray}
where $m(r,l) = m\theta(r-r_{0})+(r^{3}/2l^{2})\theta(r_{0}-r)$, $l=(\Lambda/3)^{-1/2}$ is the Hubble scale parameter, and $r_{0}=(12/\alpha)^{1/6}(2m/l)^{1/3}l$ is the radius of the false vacuum boundary
(we can choose the value of $\alpha$ as a free parameter).
Then, one can easily check that (if we choose $\alpha = 12$) the metric has the outer ($r_{+}=2m$)
and the inner horizon ($r_{-}=l$) and usually $r_{-}<r_{0}<r_{+}$ holds true as long as $l\ll m$.
If $r<r_{0}$, the metric is exactly the same as the de Sitter space. Otherwise, it exactly resembles a Schwarzschild black hole.
We can calculate the proper shell condition \cite{Frolov:1988vj}:
\begin{equation} \label{energy-momentum tensor}
S^{\mu}_{\nu} = \textrm{diag}\left(\frac{\lambda}{4\pi},0,\frac{\kappa+\lambda}{8\pi},\frac{\kappa+\lambda}{8\pi}\right),
\end{equation}
where
\begin{eqnarray}
 \kappa  & = & \frac{r_{0}}{l^{2}}\left[\left(\frac{r_{0}}{l}\right)^{2}-1\right]^{-1/2}+ \frac{m}{r_{0}^{2}}\left[\frac{2m}{r_{0}}-1\right]^{-1/2}, \\
 \lambda & = & \frac{1}{r_{0}}\left[\left(\frac{r_{0}}{l}\right)^2-1\right]^{1/2}- \frac{1}{r_{0}}\left[\frac{2m}{r_{0}}-1\right]^{1/2}.
\end{eqnarray}

This regular black hole model is free from any singularity since it violates the global hyperbolicity. So, it may suffer from mass inflation. This model assumes a thin-shell that mediates the inside false vacuum region and the outside true vacuum region, and the shell is space-like. We can prove that the space-like shell is stable under small perturbation \cite{Balbinot:1990}. As long as the shell is stable, we can construct a reasonable causal structure as the black hole forms and evaporates \cite{Frolov:1988vj,Yeom:2008qw}.

However, for realistic applications, it may pose some problems:
\begin{enumerate}
\item Initially, this model requires a false vacuum lump. Can it be generated by quantum tunneling?
\item Initially, the shell of the false vacuum lump is time-like. After the shell is trapped by an apparent horizon, it becomes space-like. How can this be dynamically possible?
\item Although a stationary solution is obtained, will the internal structure be stable, even in the presence of mass inflation?
\end{enumerate}
However, we think that our numerical study is still viable, even though there are some potential problems. In this paper, we do not want to remove all the singularities. Rather, we want to postpone the formation of the singularity. For a deeper inside region, we infer that a singularity will be formed. Our purpose is to modify the horizon dynamics using false vacuum bubbles. From this stance, each problem encountered above can be relaxed in the following ways:
\begin{enumerate}
\item Quantum fluctuations can generate a false vacuum lump during a short time \cite{Lee:1987qc}. Also, there is a consensus that a buildable bubble can be generated by unitary processes (although we cannot construct instantons) \cite{Freivogel:2005qh}. Moreover, there is an instanton solution of a small false vacuum bubble in modified gravity \cite{Kim:2010yr}. Therefore, the assumption made in order to have a false vacuum bubble should be in principle allowed.
\item The thin-shell dynamics can be questionable, whether it is time-like or space-like. However, by using numerical techniques \cite{Hansen:2009kn}, we can see clear dynamics of the thick bubble walls.
\item There will be mass inflation. However, we can postpone the curvature cutoff for mass inflation, by assuming sufficiently large number of scalar fields. Therefore, our analysis around the apparent horizon always makes sense.
\end{enumerate}

\subsection{Numerical setup}

In this subsection, we will discuss a numerical model that mimics that of Frolov, Markov, and Mukhanov.

\subsubsection{Double-null formalism}

We describe a Lagrangian with scalar fields $\Phi$, $\phi$, and the potential $V(\Phi)$:
\begin{eqnarray} \label{Lagrangian}
\mathcal{L} = - \frac{1}{2}\Phi_{;a}\Phi_{;b}g^{ab}-V(\Phi) - \frac{1}{2}\phi_{;a}\phi_{;b}g^{ab},
\end{eqnarray}
where $\Phi$ is used to make a false vacuum bubble and $\phi$ is used to make a black hole.
From this Lagrangian we can derive the equations of motion for the scalar field as:
\begin{eqnarray} \label{scalar}
\Phi_{;ab}g^{ab}-V^{'}(\Phi) &=& 0,\\
\phi_{;ab}g^{ab} &=& 0.
\end{eqnarray}
In addition, the energy-momentum tensor becomes
\begin{eqnarray} \label{energy_momentum}
T_{ab}=\Phi_{;a}\Phi_{;b}-\frac{1}{2}g_{ab}(\Phi_{;c}\Phi_{;d}g^{cd}+2V(\Phi)) + \phi_{;a}\phi_{;b} -\frac{1}{2}g_{ab} \phi_{;c}\phi_{;d}g^{cd} .
\end{eqnarray}

Now, we will describe our numerical setup. We start from the double-null coordinates (our convention is $[u,v,\theta,\varphi]$),
\begin{eqnarray} \label{double_null}
ds^{2} = -\alpha^{2}(u,v) du dv + r^{2}(u,v) d\Omega^{2},
\end{eqnarray}
assuming spherical symmetry. Here $u$ is the ingoing null direction and $v$ is the outgoing null direction.

We define the main functions as follows \cite{Hong:2008mw,Hansen:2009kn,Hwang:2010aj}: the metric function $\alpha$, the area function $r$, and the scalar fields $S \equiv \sqrt{4\pi} \Phi$ and $s \equiv \sqrt{4\pi} \phi$. We also use the following conventions: $d \equiv \alpha_{,v}/\alpha$, $h \equiv \alpha_{,u}/\alpha$, $f \equiv r_{,u}$, $g \equiv r_{,v}$, $W \equiv S_{,u}$, $Z \equiv S_{,v}$, $w \equiv s_{,u}$, $z \equiv s_{,v}$.

From the above setup, the following components can be calculated:
\begin{eqnarray}
G_{uu}&=&-\frac{2}{r} (f_{,u}-2fh), \\
G_{uv}&=&\frac{1}{2r^{2}} \left( 4 rf_{,v} + \alpha^{2} + 4fg \right), \\
G_{vv}&=&-\frac{2}{r} (g_{,v}-2gd), \\
G_{\theta\theta}&=&-4\frac{r^{2}}{\alpha^{2}} \left(d_{,u}+\frac{f_{,v}}{r}\right),
\end{eqnarray}
\begin{eqnarray}
T_{uu}&=&\frac{1}{4\pi} \left(W^{2} + w^{2}\right), \\
T_{uv}&=&\frac{\alpha^{2}}{2} V(S), \\
T_{vv}&=&\frac{1}{4\pi} \left(Z^{2} + z^{2}\right), \\
T_{\theta\theta} &=& \frac{r^{2}}{2\pi\alpha^{2}} \left( WZ + wz \right) - r^{2} V(S),
\end{eqnarray}
where
\begin{eqnarray}
V(S) = V(\Phi) |_{\Phi = S/\sqrt{4\pi}}.
\end{eqnarray}

From the equations of motion of the scalar fields, we get the following conditions:
\begin{eqnarray} \label{scalar_2}
rZ_{,u}+fZ+gW+ \pi \alpha^{2}rV^{'}(S) &=&0,\\
rz_{,u}+fz+gw &=&0.
\end{eqnarray}
Note that, $V^{'}(S) = dV(S)/dS$.

We also consider renormalized energy-momentum tensors to include the semiclassical effects. The spherical symmetry makes it reasonable to use the $1+1$-dimensional results \cite{Davies:1976ei,Birrell:1982ix} divided by $4\pi r^{2}$ \cite{Hong:2008mw,doublenull,Hansen:2009kn,Hwang:2010aj}:
\begin{eqnarray}
\langle \hat{T}_{uu} \rangle &=& \frac{P}{4\pi r^{2}}\left(h_{,u}-h^{2}\right) \label{Tq1},
 \\
\langle \hat{T}_{uv} \rangle = \langle \hat{T}_{vu} \rangle &=& -\frac{P}{4\pi r^{2}}d_{,u} \label{Tq2},
 \\
\langle \hat{T}_{vv} \rangle &=& \frac{P}{4\pi r^{2}}\left(d_{,v}-d^{2}\right) \label{Tq3},
\end{eqnarray}
with $P \equiv Nl_{\mathrm{Pl}}^2 / 12\pi$, where $N$ is the number of massless scalar fields and $l_{\mathrm{Pl}}$ is the Planck length.
We use the semi-classical Einstein equation,
\begin{eqnarray}
G_{\mu\nu}=8\pi \left( T_{\mu\nu}+\langle \hat{T}_{\mu\nu} \rangle \right).
\end{eqnarray}

Finally, we summarize our simulation equations:
\begin{enumerate}
\item \emph{Einstein equations:}
\begin{eqnarray}
d_{,u} = h_{,v} &=& \frac{1}{1-\frac{P}{r^{2}}} \left[ \frac{fg}{r^{2}} + \frac{\alpha^2}{4r^{2}} - \left(WZ +wz\right) \right], \label{sing} \\
g_{,v} &=& 2dg - r \left(Z^{2} + z^{2}\right) - \frac{P}{r}(d_{,v}-d^{2}), \label{rvv}\\
g_{,u} = f_{,v} &=& -\frac{fg}{r} - \frac{\alpha^{2}}{4r} + 2\pi\alpha^2 r V(S) - \frac{P}{r}d_{,u}, \label{ruv}\\
f_{,u} &=& 2fh - r \left(W^{2} + w^{2} \right) - \frac{P}{r}(h_{,u}-h^{2}) \label{ruu}.
\end{eqnarray}
\item \emph{Scalar field equations:}
\begin{eqnarray}
Z_{,u} = W_{,v} &=& - \frac{fZ}{r} - \frac{gW}{r} - \pi \alpha^{2}V^{'}(S),\\
z_{,u} = w_{,v} &=& - \frac{fz}{r} - \frac{gw}{r}.
\end{eqnarray}
\end{enumerate}

\subsubsection{Initial conditions and integration schemes}

We prepare a false vacuum bubble along the initial ingoing surface. We need initial conditions for each function on initial $u=u_{\mathrm{i}}=0$ and $v=v_{\mathrm{i}}=0$ surfaces.
There are two kinds of information: geometry ($\alpha, r, g, f, h, d$) and matter ($S, W, Z, s, w, z$).

On the geometry side, we have the gauge freedom to choose the initial $\alpha$ function and integrate $r$ using equations. We choose $\alpha(0,0)=\alpha_{0}$ and $h(u,0)=d(0,v)=0$. $\alpha_{0}$ is related to the mass function: $m(u,v)=(r/2)(1+4fg/\alpha^{2})$. For a fixed $r(0,0)=10$, $f(0,0)=-1/2$, and $g(0,0)=1/2$, to satisfy $m(0,0)=0$, $\alpha(0,0)=1$ is determined automatically.

On the matter side, we fix $s(u,v_{\mathrm{i}})=0$ and
\begin{eqnarray}
s(u_{\mathrm{i}},v)=A \sin^{2} \left( \pi \frac{v-v_{\mathrm{i}}}{\delta v} \right)
\end{eqnarray}
for $v_{\mathrm{i}} \leq v < v_{\mathrm{i}}+\delta v$ and otherwise $s(0,v)=0$. Then, one can calculate $s(u,v_{\mathrm{i}})$, $w(u,v_{\mathrm{i}})=s_{,u}(u,v_{\mathrm{i}})$, $s(u_{\mathrm{i}},v)$, and $z(u_{\mathrm{i}},v)=s_{,v}(u_{\mathrm{i}},v)$; in addition, $z(u,v_{\mathrm{i}})$ and $w(u_{\mathrm{i}},v)$ are obtained using the equation for $s_{,uv}$.
Additionally, we fix $S(u_{\mathrm{i}},v)=0$ and
\begin{eqnarray}
S(u,v_{\mathrm{i}}) = \left\{ \begin{array}{ll}
0 & u < u_{\mathrm{shell}},\\
\Delta \sin^{2} \left[\frac{\pi(u-u_{\mathrm{shell}})}{2\delta u_{\mathrm{shell}}}\right] & u_{\mathrm{shell}} \leq u < u_{\mathrm{shell}}+\delta u_{\mathrm{shell}},\\
\Delta & u_{\mathrm{shell}}+\delta u_{\mathrm{shell}} \leq u.
\end{array} \right.
\end{eqnarray}
Then, one can calculate $S(u,v_{\mathrm{i}})$, $W(u,v_{\mathrm{i}})=S_{,u}(u,v_{\mathrm{i}})$, $S(u_{\mathrm{i}},v)$, and $Z(u_{\mathrm{i}},v)=S_{,v}(u_{\mathrm{i}},v)$; in addition, $Z(u,v_{\mathrm{i}})$ and $W(u_{\mathrm{i}},v)$ is obtained using the equation for $S_{,uv}$.
The potential is free to choose and we fix a simple form:
\begin{eqnarray}
V(S) = V_{\mathrm{fv}} \left[ B \left(\frac{S}{\Delta} \right)^{4} - 2 \left(B+1\right) \left(\frac{S}{\Delta} \right)^{3} + \left(B+3\right) \left(\frac{S}{\Delta} \right)^{2} \right],
\end{eqnarray}
so that it has the true vacuum at $S=0$ and $V(0)=0$ while it has the false vacuum at $S=\Delta$ and $V(\Delta)=V_{\mathrm{fv}}$.

Then, as one fixes $s, w, z$ and $S, W, Z$ for initial surfaces, one can obtain $g$ and $f$ by integrating Einstein equations. And, finally, $r$ can be obtained by integrating $g$ and $f$. This finishes the assignments of the initial conditions. We observed the convergence and the consistency of the simulations in Appendix. Here, we used the 2nd order Runge-Kutta method \cite{nr}.

We fix the free parameters as follows: $r_{0}=10$, $\alpha_{0}=1$, $A=0.025$, $\delta v = 0.2$, $\delta u_{\mathrm{shell}} = 0.005$, $u_{\mathrm{shell}}=13.846$, $B=10$, $V_{\mathrm{fv}}=0.005$, $P=0.1$. The only remaining parameter to be fixed is the field value of the false vacuum $\Delta$. Of course, in general, the other parameters are free to be chosen as one likes, in principle.

\subsection{Causal structure}

\begin{figure}
\begin{center}
\includegraphics[scale=0.5]{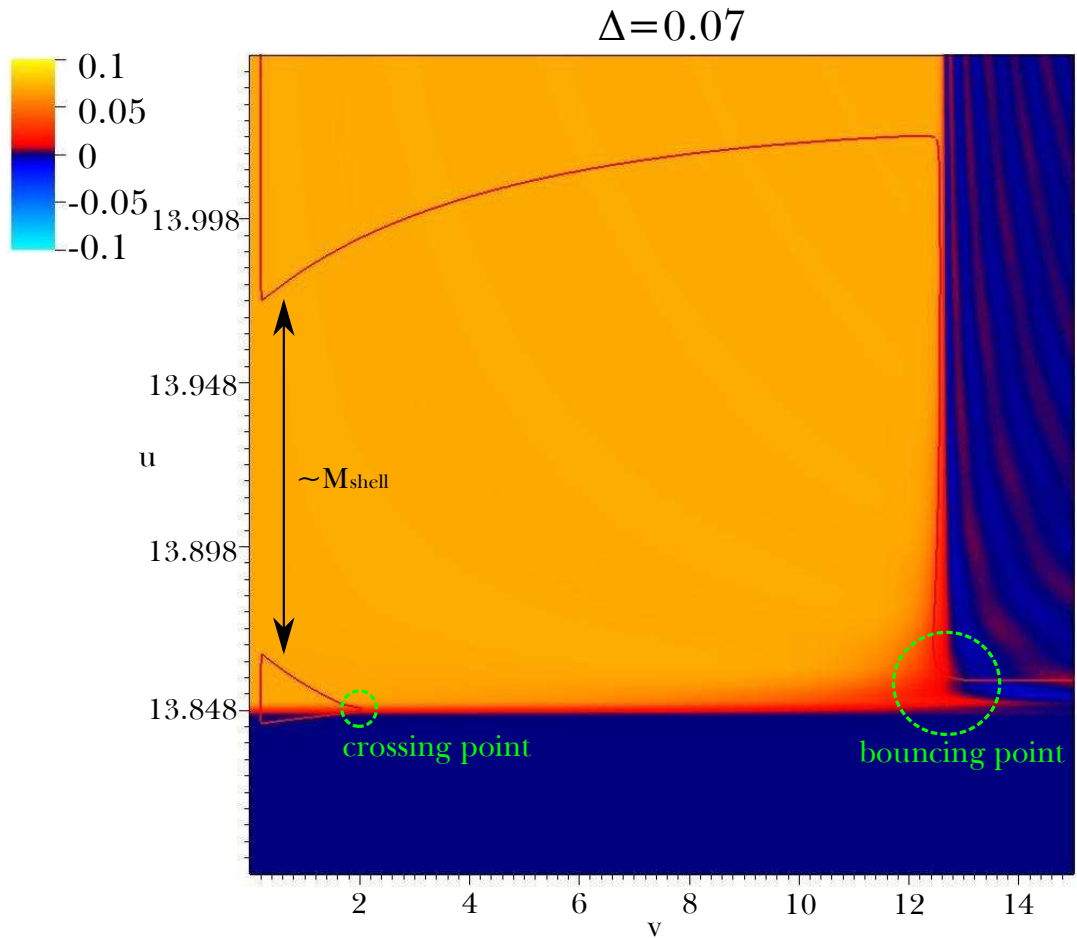}
\caption{\label{fig:G007}General global structure of a gravitational collapse in the presence of a false vacuum bubble. Here, $\Delta=0.07$. Color denotes the field value $S$ and red curves denote the apparent horizons.}
\end{center}
\end{figure}

\begin{figure}
\begin{center}
\includegraphics[scale=0.9]{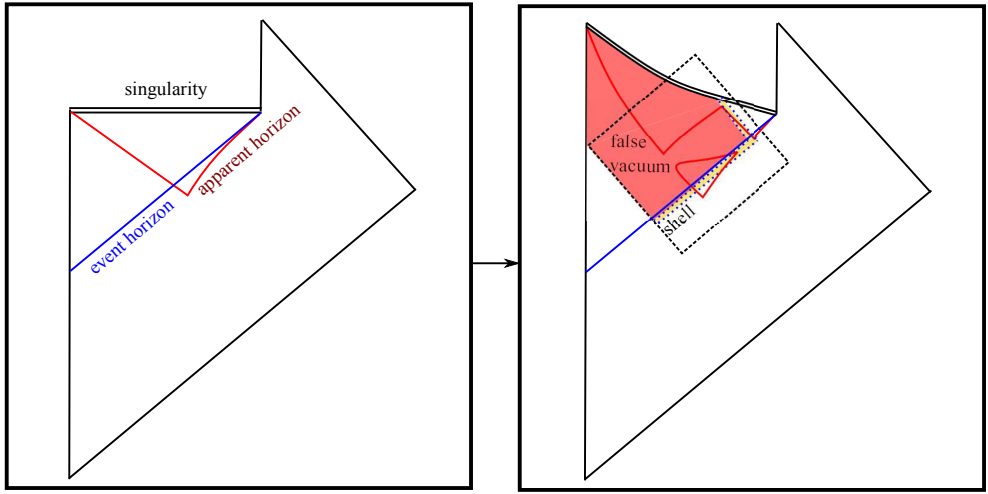}
\caption{\label{fig:singularity_2}Left: The causal structure of an evaporating neutral black hole. Right: An out-going false vacuum bubble changes the singularity structure. Red curve denotes an apparent horizon, blue curve is an event horizon, yellow region is a shell, red region depicts a false vacuum region, and dashed rectangular region is the integration domain of our simulations.}
\end{center}
\end{figure}

Figures~\ref{fig:G007} and \ref{fig:singularity_2} represent a general global causal structure of the gravitational collapse with a false vacuum bubble. It is interesting to note that this causal structure is qualitatively same as the causal structure found in \cite{Yeom:2008qw}. There are few remarkable structures. First, there appears a closed trapped region: the outer part being an outer apparent horizon and the inner part being an inner apparent horizon. Two horizons emerge at a certain \textit{crossing point}. The crossing point appears since the false vacuum shell crosses there. The out-going false vacuum shell eventually turns back to the in-going direction and we call this \textit{bouncing point}. Before a bouncing point appears, there can be an apparent horizon inside the false vacuum region. If the final black hole mass is order of $\sim M$, then the apparent horizon inside the false vacuum region is of the order of $\sim M - M_{\mathrm{shell}}$ and it swallows the shell mass $\sim M_{\mathrm{shell}}$ later. Therefore, the approximate difference in radius between a crossing point and the inside apparent horizon should be of the order of the false vacuum shell mass $\sim M_{\mathrm{shell}}$.

\begin{figure}
\begin{center}
\includegraphics[scale=0.25]{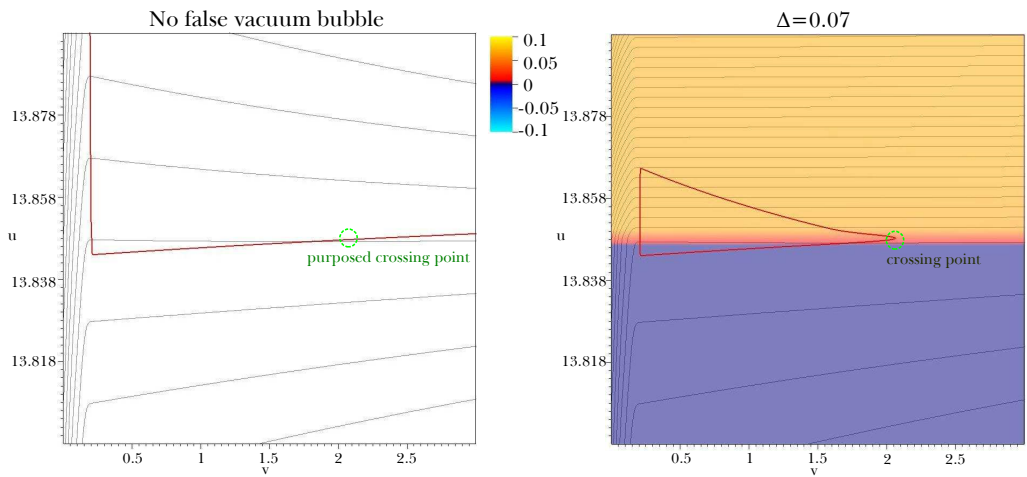}
\caption{\label{fig:crossing}Construction of the crossing point. Black contours denote $r$ and the difference between each contour is $0.01$. Color denotes $S$. Left is the case where there is no false vacuum bubble and Right is the case when $\Delta = 0.07$.}
\end{center}
\end{figure}

\begin{figure}
\begin{center}
\includegraphics[scale=0.25]{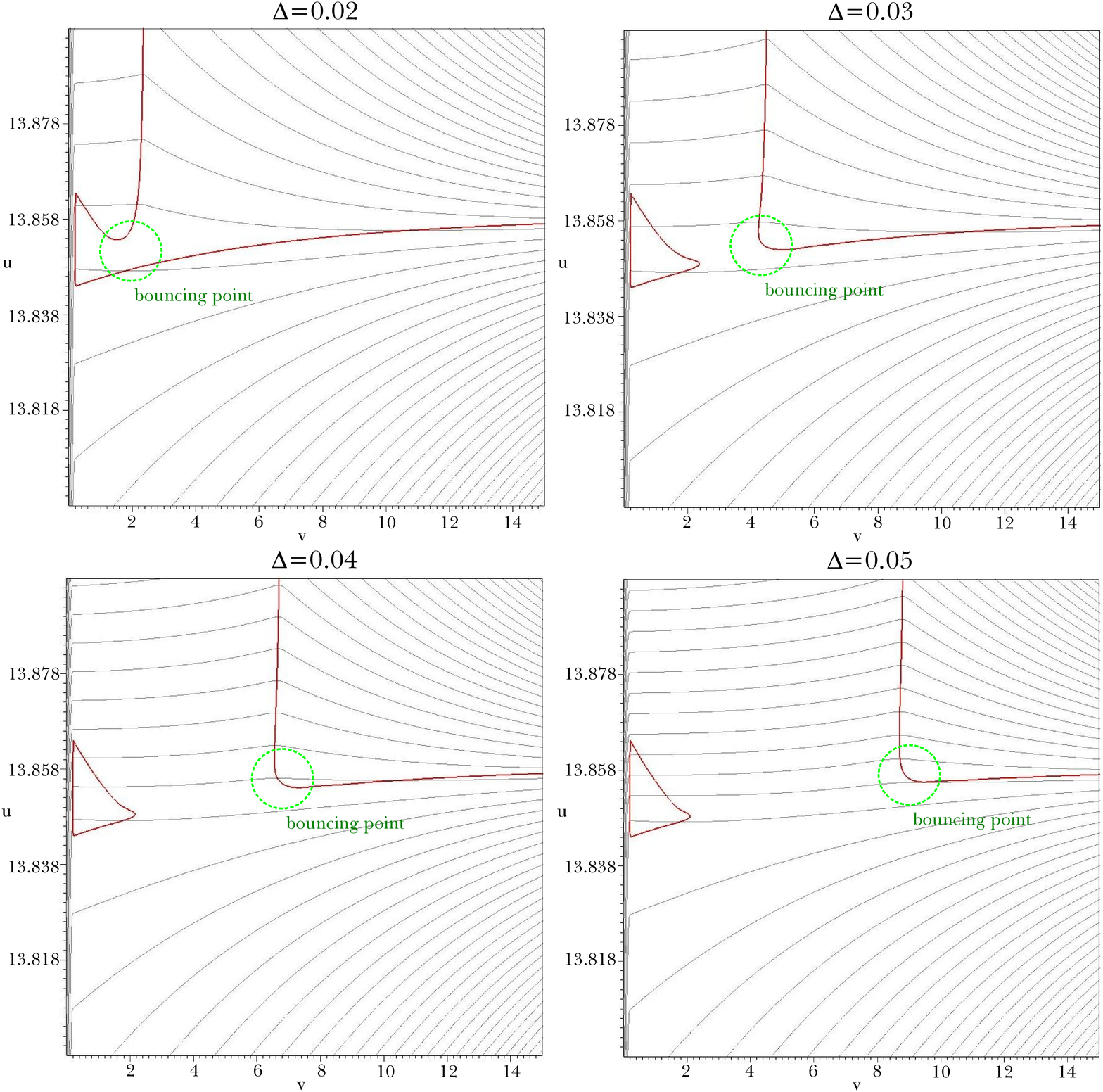}
\caption{\label{fig:shift}Shifting of the bouncing points. Black contours denote $r$ and the difference between each contour is $0.01$.}
\end{center}
\end{figure}

Now we will discuss details of the crossing point and the bouncing point. First, Figure~\ref{fig:crossing} compares two cases, black hole with and without a false vacuum bubble. For an evaporating black hole, one can relocate the crossing point to any place on the time-like horizon. This crossing point can be chosen around the information retention time, in principle. By tuning the initial location of the shell $u_{\mathrm{shell}}$ and the thickness of the shell $\delta u_{\mathrm{shell}}$ (this should be sufficiently thin), one can shoot the out-going shell to hit the desired crossing point (Figure~\ref{fig:crossing}). For this, we require that the shell energy to be sufficiently large so that the bouncing point arises sufficiently far from the crossing point and the shell does not collapse before it reaches the crossing point. Figure~\ref{fig:shift} shows such a behavior; as the shell energy increases by tuning the field amplitude $\Delta$, the bouncing point shifts to the right side, while the crossing point is not significantly changed.

In summary, we have three free parameters that depends only on the potential and the initial conditions: $u_{\mathrm{shell}}$, $\delta u_{\mathrm{shell}}$, and $\Delta$. First, we choose a specific crossing point (for our purposes, around the information retention time). Then, one can estimate the required $u_{\mathrm{shell}}$ and $\delta u_{\mathrm{shell}}$. However, these choices cannot make sure as the shell hits the crossing point if it can collapse too quickly. Now, we have to tune $\Delta$ to set a proper tension of the shell so that the bouncing point arises sufficiently far from the desirable crossing point. In this limit, the shell energy $M_{\mathrm{shell}}$ is essentially determined in terms of $\delta u_{\mathrm{shell}}$ and $\Delta$. It is easy to choose sufficiently large $M_{\mathrm{shell}}$ so that the distance between the crossing point and internal horizon structures becomes sufficiently large. Between the crossing point and the internal horizon structures, we do not have any evidence of mass inflation, although this can be observed from the deeper inside of an expanding false vacuum bubble.

\subsection{Gedanken experiments}

\subsubsection{Duplication experiments}

In this causal structure (Figure~\ref{fig:singularity_2}), the duplication experiment is well-defined and possible (Figure~\ref{fig:wall_2}). We use the same convention that was used in Section~\ref{sec:black}. Here, we can assume that the crossing point is around the information retention time. To see the duplication, the message can be sent after $\Delta t \sim M_{\mathrm{shell}}$ and this can be reasonably large. If there is mass inflation, then this can be regularized by assuming a large number of scalar fields, although there is no numerical evidence of mass inflation.

\subsubsection{Where is the firewall?}

\begin{figure}
\begin{center}
\includegraphics[scale=0.7]{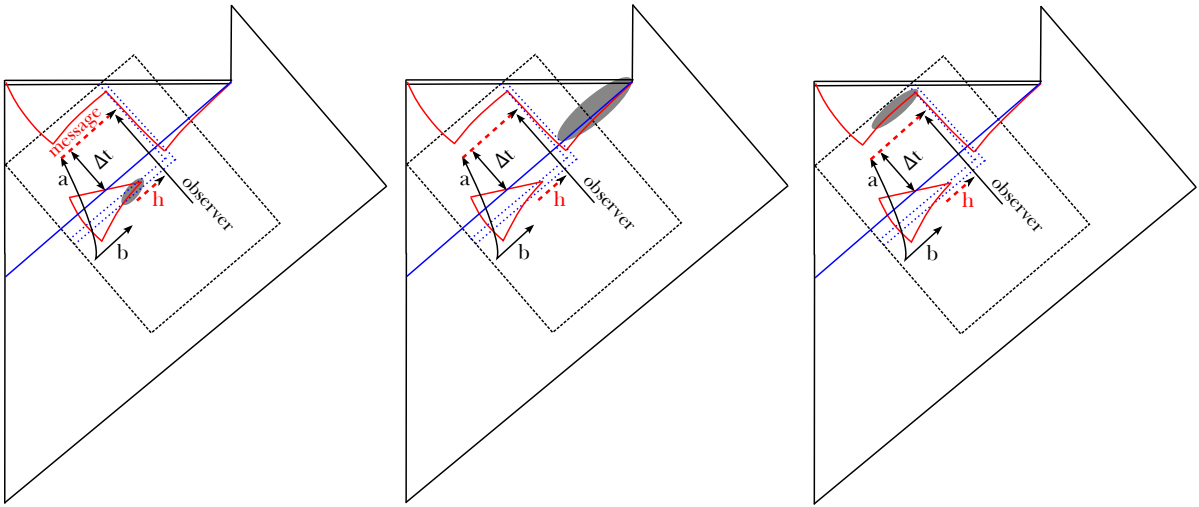}
\caption{\label{fig:wall_2}Left: To resolve the inconsistency of the black hole complementarity, the firewall should exist on the outer apparent horizon. Middle and Right: If the firewall exists on some other place, then it cannot cure the inconsistency due to black hole complementarity.}
\end{center}
\end{figure}

Therefore, it is inevitable to introduce the firewall, if one wants to maintain the black hole complementarity.
As was discussed in Section~\ref{sec:black}, the firewall should be located on the apparent horizon. Note that this situation is quite different from that of the Schwarzschild black hole, since the apparent horizon is \textit{disconnected}. Figure~\ref{fig:wall_2} shows possible candidates for the firewall. If the firewall is on the internal horizon structures (Middle and Right of Figure~\ref{fig:wall_2}), then it cannot resolve the problem of black hole complementarity. Therefore, it should be located on the outer apparent horizon after the information retention time. However, we know that the horizon suddenly disappears.

Here, our first question is this: \textit{what will be observed by the in-falling duplication observer?} The answer is simple. If the firewall can kill the message generated by $a$, then the observer can notice the effect of the firewall and we infer that the firewall can affect the causal future of the spacetime.

Note that obviously the firewall is situated outside the event horizon. If we accept that the firewall can affect the causal future, then the question that comes next is: \textit{does the firewall affect to the future infinity?} If the apparent horizon is connected, then one can believe that any effect can be screened by the apparent horizon. However, if the apparent horizon is not connected, there is no consistent screen. It is fair to say that the \textit{firewall is now naked}. Therefore, we conclude that the firewall should affect the asymptotic future infinity.

Therefore, it reveals a paradox if we assume the following two contents: the firewall should prevent the duplication experiment and should not affect the future infinity. From the former assumption, the firewall should be on the apparent horizon. Therefore, if the firewall is a kind of singularity, then \textit{it should be a time-like singularity}, since the outer apparent horizon is time-like. If we consider the Schwarzschild black hole, one could ignore this problem, since the apparent horizon is \textit{connected}, and hence the inside and the outside of the firewall is well-separated. However, if we consider \textit{disconnected} apparent horizons, it is clear that one cannot ignore the effects due to the firewall along the out-going direction (Figure~\ref{fig:wall_3}). This somehow requires quantum gravitational modification of the outside of a black hole. \textit{A firewall cannot prevent the modification of semi-classical quantum field theory.}

\subsubsection{Violation of cosmic censorship?}

\begin{figure}
\begin{center}
\includegraphics[scale=0.7]{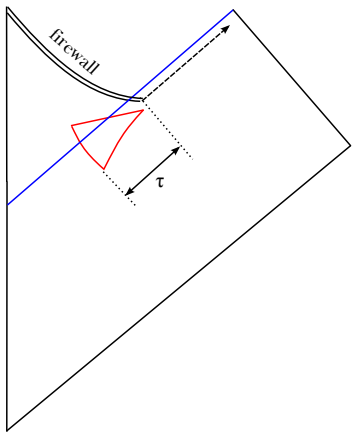}
\caption{\label{fig:wall_4}If the crossing point is slightly before the information retention time, during this time, a firewall grows and approaches the apparent horizon. It is also possible that a firewall is outside the event horizon while the apparent horizon gets disappeared. Then the effect of the firewall should be observed by an asymptotic observer.}
\end{center}
\end{figure}

Furthermore, now let us consider the situation when the crossing point is slightly before the information retention time (Figure~\ref{fig:wall_4}). During the time evolution, the firewall grows and approaches the apparent horizon \cite{Susskind:2012rm}. We can tune this in such a way that the crossing point and the information retention time are quite close to each other so that the firewall can grow even outside the event horizon. Then, there is no way to screen the effect of the firewall-singularity. Again, the firewall cannot prevent the modification of the semi-classical quantum field theory for an asymptotic observer. Then, is it a kind of violation of strong cosmic censorship?

\subsubsection{Conclusion}

In conclusion, the firewall was originally a conservative way to maintain the black hole complementarity. To prevent the duplication experiment, the firewall should be very close to the apparent horizon, and hence, this should be outside the event horizon. We asked whether the firewall-singularity that is outside the event horizon must affect the future infinity or can be screened by the apparent horizon; the latter is a stronger assumption to help the original version of the firewall proposal. However, our numerical simulations reveal the fact that the firewall should affect the future infinity and asymptotic observer, even with our stronger assumption.

Therefore, we give some possible interpretations:
\begin{enumerate}
\item We can have more \textit{ad hoc} assumptions in order to maintain the firewall idea -- so that, we only modify Assumption~$5$. For example, the Horowitz-Maldacena proposal \cite{Horowitz:2003he} can be potentially relevant, although the idea in itself has other potential problems \cite{Gottesman:2003up} and can be falsified with large $N$ rescaling \cite{Hong:2008ga}.
\item We may accept that it is inevitable to assume macroscopic effects of quantum gravity, even for a semi-classical system (Assumption $4$ should be modified) \cite{Giddings:2001pt}. This possibility can be related to the fuzzball interpretation \cite{Mathur:2012jk,Chowdhury:2012vd,Bena:2012zi}.
\item We may think that one of the Assumptions $1$, $2$ or $3$ should be modified \cite{Yeom:2009zp}.
\end{enumerate}

\section{\label{sec:dis}Discussion}

In this paper, we have discussed the black hole complementarity, the firewall proposal, and related gedanken experiments.

We have illustrated five important assumptions: unitarity, entropy-area formula, existence of an information observer, semi-classical quantum field theory for an asymptotic observer, and general relativity for an in-falling observer. These five assumptions require a duplication of information around the event horizon and hence the black hole complementarity for consistency of the theory. However, if there is an observer who can see the duplication of information, then the black hole complementarity can be falsified.

Black hole complementarity is indeed falsified with the two arguments: large $N$ rescaling and the AMPS argument. Especially, the former is useful to show how communication happens between two observers: asymptotic and in-falling. To resolve this contradiction, AMPS introduced the firewall. If a firewall prevents the duplication experiment, then it should be located close to the apparent horizon after a certain time scale (information retention time or scrambling time).

We also tried to find out whether these two assumptions are consistent at the same time: (1) the firewall around the apparent horizon and (2) the firewall only affecting inside the black hole. To check the consistency, in this paper, we have considered a gravitational collapse with a false vacuum lump, which is motivated by a regular black hole model. We were able to construct an example where the apparent horizon can be disconnected. Then, one can clearly see that there are no barrier to screen the out-going effects due to the firewall and the firewall can be naked.

From these arguments, we can clearly conclude the following:
\begin{description}
\item[Conclusion 1:] The original version of the black hole complementarity is inconsistent. An in-falling observer and an asymptotic observer can communicate.
\item[Conclusion 2:] To keep the basic philosophy of black hole complementarity, we need a firewall outside the event horizon to kill the information of the in-falling information.\footnote{Of course, if we do not trust the philosophy of the black hole complementarity, then there is no need for a firewall.}
\item[Conclusion 3:] A consistent firewall should affect not only an in-falling observer, but also an asymptotic observer.
\end{description}

Therefore, there may be three possibilities: we need more assumptions to maintain the black hole complementarity in \textit{ad hoc} ways, we have to take care of the macroscopic effects due to quantum gravity, or we have to modify the traditional entropy-area formula, etc. In this paper, we are not in a position to judge which is the final answer and this is postponed for a future work.

\section*{Appendix: Consistency and convergence tests}

In this appendix, we report on the convergence and consistency tests for our simulations. As a demonstration, we consider the case with $\Delta = 0.07$.

\begin{figure}
\begin{center}
\includegraphics[scale=1]{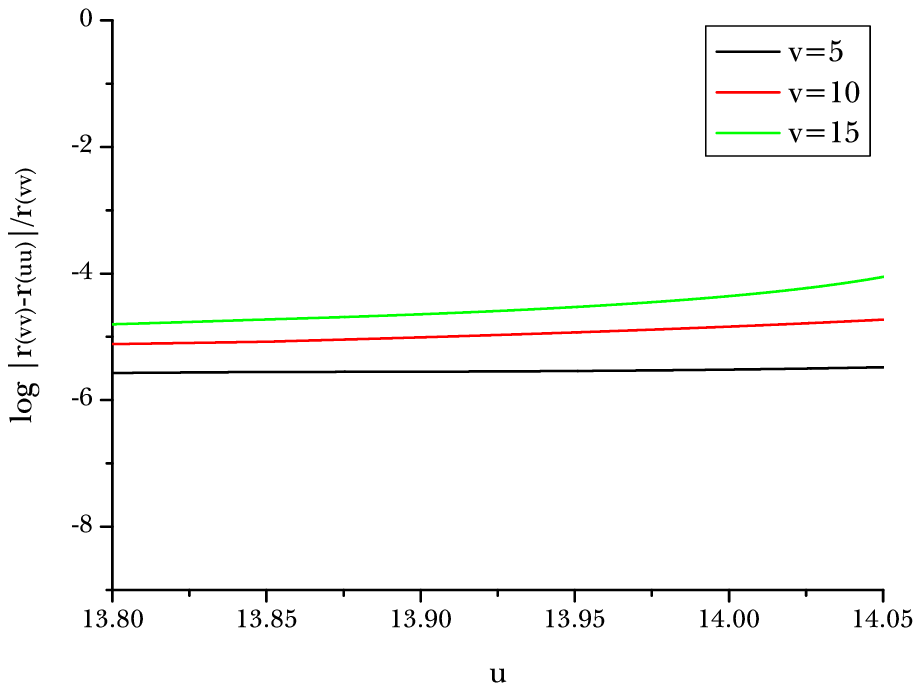}
\caption{\label{fig:constraint}$\log |r_{(vv)}-r_{(uu)}|/r_{(vv)}$ for $\Delta = 0.07$.}
\includegraphics[scale=1]{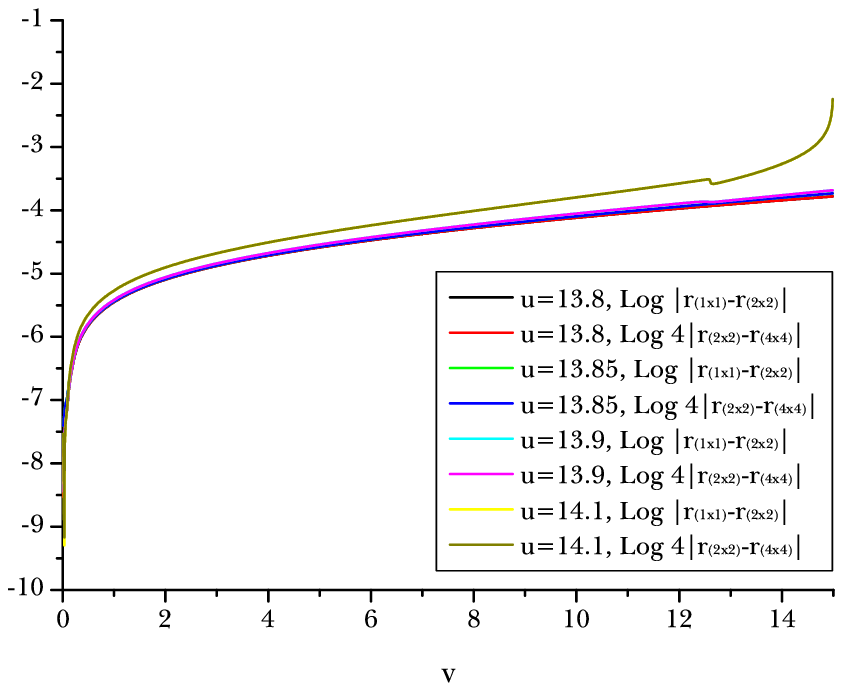}
\caption{\label{fig:convergence}Convergence tests for $\Delta = 0.07$.}
\end{center}
\end{figure}

For consistency, we compare $r$ with different schemes. In this paper, we obtain $r$ by integrating $r_{,v}$, where $r_{,v}$ is obtained by integrating the equation for $r_{,vv}$: we call this $r_{(vv)}$. However, this is not the unique choice. We can obtain $r$ by integrating $r_{,u}$, where $r_{,u}$ is obtained by integrating the equation for $r_{,uu}$: we call this $r_{(uu)}$. In Figure~\ref{fig:constraint} shows $\log |r_{(vv)}-r_{(uu)}|/r_{(vv)}$ around $v=5, 10, 15$. The difference of two schemes are less than $10^{-4}$. This shows that the error from the constraint equations (equations that we did not use for numerical integration) is sufficiently small and not accumulated. Therefore, this shows a good consistency.

For convergence, we compared finer simulations: $1\times1$, $2\times2$, and $4\times4$ times finer around $u=13.8, 13.85, 13.9, 14.1$. In Figure~\ref{fig:convergence}, we see that the difference between the $1\times1$ and $2\times2$ times finer cases is $4$ times the difference between the $2\times2$ and $4\times4$ times finer cases, and thus our simulation converges to second order. The numerical error is $\lesssim 10^{-3}$ except for the region near the singularity.

\section*{Acknowledgment}

DY would like to thank Hanno Sahlmann for the hospitality and helpful discussions during a visit at Erlangen-N\"{u}rnberg University. The authors would like to thank for helpful comments of Borun Chowdhury, Andrea Puhm, and a referee of JCAP to point out some issues. DY, DH and BHL are supported by the National Research Foundation of Korea(NRF) grant funded by the Korea government(MEST) through the Center for Quantum Spacetime(CQUeST) of Sogang University with grant number 2005-0049409. DH is supported by Korea Research Foundation grants (KRF-313-2007-C00164, KRF-341-2007-C00010) funded by the Korean government (MOEHRD) and BK21.

\end{document}